# Electromagnetic angular momentum of quantized wavepackets in free space

Masud Mansuripur
James C. Wyant College of Optical Sciences, The University of Arizona, Tucson

**Abstract**. A single electromagnetic plane-wave propagating in free space possesses neither spin nor orbital angular momentum. Both types of angular momentum arise from interference between pairs of plane-waves having the same temporal frequency $\omega$ but differing $k$-vectors $\boldsymbol{k}_1$ and $\boldsymbol{k}_2$. While it is fairly straightforward to evaluate a wavepacket's spin and orbital angular momenta in the $(\boldsymbol{k},\omega)$ continuum by means of Fourier transformation, obtaining the same results by discretizing the $(\boldsymbol{k},\omega)$ space, then attempting to approach the continuum limit via an infinite enlargement of the spatial volume under consideration, is fraught with danger.

**1. Introduction**. In classical as well as quantum electrodynamics, evaluating the optical angular momentum of a wavepacket in free space requires careful consideration of the multimodal nature of the wavepacket.[1-4] This paper addresses some of the nuances of a numerical (i.e., discrete) method of computation, where the separation of spin from orbital angular momentum is reexamined and attention is drawn to certain features of the classical as well as quantum-optical energy and angular momenta of free wavepackets. Analytical results for an electromagnetic (EM) wavepacket in free space possessing a continuous plane-wave spectrum have been previously reported[3] using transformation from the usual spacetime domain $(\boldsymbol{r},t)=(x,y,z,t)$ to the Fourier domain $(k_x,k_y,\omega)$. There also exists a more general analysis which applies to EM waves in the presence of a number of electric point-charges in arbitrary motion in the $(\boldsymbol{r},t)$ domain, where the Fourier transformation is done over the space coordinates only, yielding expressions for the angular momenta in the form of integrals over the $(k_x,k_y,k_z)$ domain.[1] In the special case of free EM waves (i.e., absent all electric charges and currents), the results reported in Ref.[3] are in complete accord with those of Ref.[1], as demonstrated in Ref.[4]. The concern of the present paper is the computation of the energy and angular momentum content of a discrete set of plane-waves in free space, with an eye toward the similarities and differences with the continuum results reported in Refs.[1-4].

We begin in Sec.2 by computing the EM energy and angular momentum for a wavepacket consisting of a discrete set of propagating plane-waves in free space. The discretization is done on a dense rectangular mesh in the $k$-space, and we obtain expressions for both the spin and orbital angular momenta of the wavepacket. It is observed that the energy content of the wavepacket is a sum over the energies of its constituent plane-waves, whereas both spin and orbital angular momenta require contributions not only from each plane-wave but also from its six nearest neighbors on the rectangular mesh in the $k$-space. The results of Sec.2 are nearly identical with those reported for a continuum of plane-waves,[1,3] although the angular momenta turn out to be twice as large as those of a wavepacket with a continuous plane-wave spectrum. In the remaining part of the paper we attempt to understand and explain the origins of this unwelcome factor of 2.

Section 3 examines a pair of propagating plane-waves in free space inside a hypothetical spherical volume $V$ in the limit when $V\to\infty$. When the $k$-vectors $\boldsymbol{k}_1$ and $\boldsymbol{k}_2$ of the two plane-waves approach one another, the spin and orbital angular momenta of the superposed pair exhibit the general characteristics expected from a finite wavepacket possessing a continuous plane-wave spectrum.[1,3] The only discrepancy is that the overall magnitude of the pair's total angular momentum falls well below that expected from a genuine wavepacket. Another difficulty involves the energy content of the plane-wave pair, which, in the desired $\boldsymbol{k}_1 \to \boldsymbol{k}_2$ limit, exceeds (by a factor of 2) the sum of the energies of the individual plane-waves inside the volume $V$. In spite of the above problems, this simple example brings us tantalizingly close to understanding the essence of optical angular momentum (both spin and orbital) as being rooted in interference between pairs of nearly identical plane-waves. The bottleneck appears to be in striking the right balance between the spatial volume $V$ under consideration and the "proximity" of the $k$-vectors of the plane-wave pair.

Suspecting that the difficulties encountered in Sec.3 stem from the fact that the spherical volume $V$ harbors only a small fraction of a single interference fringe in the appropriate $\boldsymbol{k}_1 \to \boldsymbol{k}_2$ limit, we consider in Sec.4 the energy and the angular momentum of a pair of plane-waves contained within a cubical volume $V$ that is deliberately chosen to house a full interference fringe. The energy content of the volume is now found to be the sum of the EM energies of the two plane-waves, as anticipated. The spin and orbital angular momenta of the pair also exhibit many of the desired features of angular momentum in the continuum limit. The difficulty, as before, is with the overall magnitudes of the spin and orbital momenta, which are off by a factor of 2 compared to what one expects from a wavepacket in the continuum limit.[1,3] A secondary point of concern is that, in the presence of multiple plane-waves, each one tends to interfere with all the others of the same frequency $\omega$, thus adding to each plane-wave's share of the angular momentum without regard for the relative proximity of the participating pair of $k$-vectors. While this unwelcome side-effect of the chosen methodology seems to violate the well-established properties of wavepackets possessing a continuous plane-wave spectrum, it should *not* be of any real concern since, on closer inspection, it is revealed to be an artifact of the specific setup conjured for the arguments made in Sec.4. In a nutshell, this is because the volume $V$ that houses a single fringe (or, more generally, an integer number of fringes) for one pair of $k$-vectors does not necessarily coincide with the volume $V$ corresponding to a different pair.

The discrepancy of the results obtained for wavepackets with discrete versus continuous spectra is finally resolved in Sec.5, where we argue that the limit operation in which the volume $V$ of the wavepacket is made to approach infinity must take precedence over the summation operation in which the contributions of the various plane-waves are combined. With the subtle mathematical consequences of $V \to \infty$ properly understood and subsequently taken under consideration, the ambiguities surrounding the "proximity" of each $k$-vector to its neighbors on a discrete mesh in the $k$-space will be clarified. The paper closes with a few concluding remarks in Sec.6.

**2. Electromagnetic energy and angular momentum in a discretized Cartesian $\boldsymbol{k}$-space**. At a fundamental level, the EM angular momentum is produced by pairs of propagating (and interfering) plane-waves in free space. Taking a very large $L \times L \times L$ cube in the three-dimensional Euclidean space identified by the $xyz$ Cartesian coordinates, we assume that the allowed (discretized) wave-vectors are $\boldsymbol{k}^{(n)} = (\omega^{(n)}/c)\hat{\boldsymbol{\kappa}}^{(n)} = (2\pi/L)(n_x\hat{\boldsymbol{x}} + n_y\hat{\boldsymbol{y}} + n_z\hat{\boldsymbol{z}})$, with $\omega^{(n)} \geq 0$ and $\boldsymbol{n} = (n_x, n_y, n_z)$ being a triplet of positive, zero, or negative integers. The individual plane-waves occupying the region $\boldsymbol{r} = (x, y, z) \in [-L/2, L/2]^3$ are specified in terms of their $\boldsymbol{E}$ and $\boldsymbol{B}$ fields, as follows:†

$$\boldsymbol{E}^{(n)}(\boldsymbol{r}, t) = \mathrm{Re}\{\boldsymbol{E}_0^{(n)} \exp[\mathrm{i}(\boldsymbol{k}^{(n)} \cdot \boldsymbol{r} - \omega^{(n)} t)]\}, \tag{1}$$

$$\boldsymbol{B}^{(n)}(\boldsymbol{r}, t) = \mathrm{Re}\{\boldsymbol{B}_0^{(n)} \exp[\mathrm{i}(\boldsymbol{k}^{(n)} \cdot \boldsymbol{r} - \omega^{(n)} t)]\}. \tag{2}$$

† Choosing the center of the $L \times L \times L$ cube at the origin of the coordinate system simplifies subsequent calculations.

The $\boldsymbol{E}$ and $\boldsymbol{B}$ field amplitudes $\boldsymbol{E}_0^{(n)}$ and $\boldsymbol{B}_0^{(n)}$, their corresponding $k$-vector $\boldsymbol{k}^{(n)}$, and their frequency $\omega^{(n)}$ are interrelated through Maxwell's equations in the following way:

$$\boldsymbol{k}^{(n)} \cdot \boldsymbol{E}_0^{(n)} = 0, \tag{3}$$

$$\boldsymbol{k}^{(n)} \cdot \boldsymbol{B}_0^{(n)} = 0, \tag{4}$$

$$\boldsymbol{k}^{(n)} \times \boldsymbol{E}_0^{(n)} = \omega^{(n)} \boldsymbol{B}_0^{(n)}. \tag{5}$$

The EM angular-momentum-density is given by $\boldsymbol{j}(\boldsymbol{r},t) = \boldsymbol{r} \times [\varepsilon_0 \boldsymbol{E}(\boldsymbol{r},t) \times \boldsymbol{B}(\boldsymbol{r},t)]$. In this expression, the cross-product of $\boldsymbol{E}$ and $\boldsymbol{B}$ consists of terms whose $E$-field comes from the plane-wave having the index triplet $\boldsymbol{m}$, while the $B$-field belongs to the plane-wave whose index triplet is $\boldsymbol{n}$. At first, we shall take $\boldsymbol{m}$ and $\boldsymbol{n}$ to be arbitrary, but it will soon emerge that the $(\boldsymbol{m}, \boldsymbol{n})$ pair of plane-waves can make a nonzero contribution to the overall angular momentum only when they differ by 1 in a single element of their corresponding index triplets. For instance, the $(\boldsymbol{m}, \boldsymbol{n})$ pair will contribute to the overall angular momentum if $(n_x, n_y, n_z) = (m_x, m_y \pm 1, m_z)$, and so on. To appreciate the logic of this assertion, consider the contribution to the overall angular momentum by the cross-product of $\boldsymbol{E}^{(m)}$ and $\boldsymbol{B}^{(n)}$, namely,[‡]

$$\boldsymbol{\mathcal{J}} = \iiint_{-½L}^{½L} \boldsymbol{r} \times ½\text{Re}\{\varepsilon_0 \boldsymbol{E}_0^{(m)} \exp[\mathrm{i}(\boldsymbol{k}^{(m)} \cdot \boldsymbol{r} - \omega^{(m)} t)] \times \boldsymbol{B}_0^{(n)*} \exp[-\mathrm{i}(\boldsymbol{k}^{(n)} \cdot \boldsymbol{r} - \omega^{(n)} t)]\} \mathrm{d}x\mathrm{d}y\mathrm{d}z$$

$$= ½\varepsilon_0 \text{Re} \iiint_{-½L}^{½L} \boldsymbol{r} \times \left\{ \boldsymbol{E}_0^{(m)} \times \left( \boldsymbol{k}^{(n)} \times \boldsymbol{E}_0^{(n)*} / \omega^{(n)} \right) e^{\mathrm{i}[(\boldsymbol{k}^{(m)} - \boldsymbol{k}^{(n)}) \cdot \boldsymbol{r} - (\omega^{(m)} - \omega^{(n)}) t]} \right\} \mathrm{d}x\mathrm{d}y\mathrm{d}z$$

$\omega^{(m)}$ must equal $\omega^{(n)}$ if $\boldsymbol{\mathcal{J}}$ is to be time-independent

$$= \left(\frac{\varepsilon_0}{2\omega^{(n)}}\right) \text{Re} \iiint_{-½L}^{½L} \boldsymbol{r} \times \left\{ [(\boldsymbol{E}_0^{(m)} \cdot \boldsymbol{E}_0^{(n)*}) \boldsymbol{k}^{(n)} - (\boldsymbol{E}_0^{(m)} \cdot \boldsymbol{k}^{(n)}) \boldsymbol{E}_0^{(n)*}] e^{\mathrm{i}(\boldsymbol{k}^{(m)} - \boldsymbol{k}^{(n)}) \cdot \boldsymbol{r}} \right\} \mathrm{d}x\mathrm{d}y\mathrm{d}z$$

$$= \left(\frac{\varepsilon_0}{2\omega^{(n)}}\right) \text{Re} \iiint_{-½L}^{½L} \{[(y k_z^{(n)} - z k_y^{(n)}) \hat{\boldsymbol{x}} + (z k_x^{(n)} - x k_z^{(n)}) \hat{\boldsymbol{y}} + (x k_y^{(n)} - y k_x^{(n)}) \hat{\boldsymbol{z}}] \boldsymbol{E}_0^{(m)} \cdot \boldsymbol{E}_0^{(n)*}$$

$$- [(y E_{0z}^{(n)*} - z E_{0y}^{(n)*}) \hat{\boldsymbol{x}} + (z E_{0x}^{(n)*} - x E_{0z}^{(n)*}) \hat{\boldsymbol{y}} + (x E_{0y}^{(n)*} - y E_{0x}^{(n)*}) \hat{\boldsymbol{z}}] \boldsymbol{E}_0^{(m)} \cdot \boldsymbol{k}^{(n)}\}$$

$$\times \exp[\mathrm{i}(\boldsymbol{k}^{(m)} - \boldsymbol{k}^{(n)}) \cdot \boldsymbol{r}] \, \mathrm{d}x\mathrm{d}y\mathrm{d}z. \tag{6}$$

The integrals over the $x, y, z$ coordinates may now be evaluated with proper attention paid to the periodic boundary conditions at $x, y, z = \pm ½L$. For example, $\int_{-½L}^{½L} x \exp[\mathrm{i}(k_x^{(m)} - k_x^{(n)}) x] \, \mathrm{d}x$ vanishes if $k_x^{(m)} = k_x^{(n)}$. However, when $k_x^{(m)} \neq k_x^{(n)}$, integration by parts yields

$$\int_{-½L}^{½L} x e^{\mathrm{i}(k_x^{(m)} - k_x^{(n)}) x} \mathrm{d}x = \frac{1}{\mathrm{i}\left(k_x^{(m)} - k_x^{(n)}\right)} \left\{ x e^{\mathrm{i}(k_x^{(m)} - k_x^{(n)}) x} \Big|_{x=-½L}^{½L} - \int_{-½L}^{½L} e^{\mathrm{i}(k_x^{(m)} - k_x^{(n)}) x} \mathrm{d}x \right\}. \tag{7}$$

In the above equation, the periodic boundary condition along the $x$-axis ensures that the second term inside the curly brackets vanishes. As for the first term, the values at the upper and lower ends of the interval are $\pm(½L) e^{\pm \mathrm{i}\pi(m_x - n_x)}$, indicating that the term equals $L$ if $m_x - n_x$ is an even integer, but that it equals $-L$ if $m_x - n_x$ is an odd integer. Below, we shall argue that the only acceptable values for $m_x - n_x$ are $\pm 1$ and that, therefore, the integral in Eq.(7) ends up being equal to $\mathrm{i}L/(k_x^{(m)} - k_x^{(n)})$.

‡ On the first line of Eq.(6), the identity $\text{Re}(\boldsymbol{a} e^{-\mathrm{i}\omega_1 t}) \times \text{Re}(\boldsymbol{b} e^{-\mathrm{i}\omega_2 t}) = \text{Re}(\boldsymbol{a} e^{-\mathrm{i}\omega_1 t}) \times ½(\boldsymbol{b} e^{-\mathrm{i}\omega_2 t} + \boldsymbol{b}^* e^{\mathrm{i}\omega_2 t}) = ½\text{Re}\{(\boldsymbol{a} \times \boldsymbol{b}) e^{-\mathrm{i}(\omega_1 + \omega_2) t} + (\boldsymbol{a} \times \boldsymbol{b}^*) e^{-\mathrm{i}(\omega_1 - \omega_2) t}\}$ has been invoked. The first term is then ignored, given that $\omega_1 + \omega_2 \neq 0$ causes this term to be time-dependent. The second term will be time-independent only if $\omega_1 = \omega_2$.

Within Eq.(6), the integral in Eq.(7) is accompanied by $\int_{-½L}^{½L} \exp[\mathrm{i}(k_y^{(m)} - k_y^{(n)})y]\,\mathrm{d}y$, which is zero unless $m_y = n_y$, and also by $\int_{-½L}^{½L} \exp[\mathrm{i}(k_z^{(m)} - k_z^{(n)})z]\,\mathrm{d}z$, which vanishes unless $m_z = n_z$. It is thus seen that $\iiint_{-½L}^{½L} x \exp[\mathrm{i}(\boldsymbol{k}^{(m)} - \boldsymbol{k}^{(n)}) \cdot \boldsymbol{r}]\,\mathrm{d}x\mathrm{d}y\mathrm{d}z$ appearing in Eq.(6) will be nonzero only if $k_x^{(m)} \neq k_x^{(n)}$ while $k_y^{(m)} = k_y^{(n)}$ and $k_z^{(m)} = k_z^{(n)}$. Moreover, since we have previously argued for the necessity of $\omega^{(m)} = \omega^{(n)}$ if the angular momentum $\boldsymbol{\mathcal{J}}$ of Eq.(6) is to be time-independent, we now conclude that $k_x^{(n)}$ must be as close as possible to $k_x^{(m)}$. Consequently, only by setting $n_x = m_x \pm 1$ (while $n_y = m_y$ and $n_z = m_z$) can one ensure that $\omega^{(n)} \to \omega^{(m)}$ in the limit when $L \to \infty$ and that, under the circumstances, the three-dimensional integral under consideration assumes the nonzero value of $\mathrm{i}L^3/(k_x^{(m)} - k_x^{(n)})$. The overall contribution of the considered $(\boldsymbol{m}, \boldsymbol{n})$ pair of plane-waves to the angular momentum $\boldsymbol{\mathcal{J}}$ is thus seen to be

$\boldsymbol{E}_0^{(m)} \cdot \boldsymbol{k}^{(m)}$, which is zero according to Maxwell's 1st equation, $\boldsymbol{\nabla} \cdot \boldsymbol{E}(\boldsymbol{r}, t) = 0$, is added here for convenience

$$-\left(\frac{\varepsilon_0 L^3}{2\omega^{(n)}}\right) \mathrm{Im}\left\{\left(\frac{k_y^{(n)}\hat{\boldsymbol{z}} - k_z^{(n)}\hat{\boldsymbol{y}}}{k_x^{(m)} - k_x^{(n)}}\right) \boldsymbol{E}_0^{(m)} \cdot \boldsymbol{E}_0^{(n)*} + \left(\frac{E_{0y}^{(n)*}\hat{\boldsymbol{z}} - E_{0z}^{(n)*}\hat{\boldsymbol{y}}}{k_x^{(m)} - k_x^{(n)}}\right) \boldsymbol{E}_0^{(m)} \cdot (\boldsymbol{k}^{(m)} - \boldsymbol{k}^{(n)})\right\}$$

$$= -\left(\frac{\varepsilon_0 L^3}{2\omega^{(m)}}\right) \mathrm{Im}\left\{\left(\frac{k_y^{(m)}\hat{\boldsymbol{z}} - k_z^{(m)}\hat{\boldsymbol{y}}}{k_x^{(m)} - k_x^{(n)}}\right) \boldsymbol{E}_0^{(m)} \cdot \boldsymbol{E}_0^{(n)*} + \left(E_{0x}^{(m)} E_{0y}^{(n)*}\hat{\boldsymbol{z}} - E_{0x}^{(m)} E_{0z}^{(n)*}\hat{\boldsymbol{y}}\right)\right\}. \tag{8}$$

The remaining integrals in Eq.(6) can be similarly evaluated, leading to

$$\begin{aligned}\boldsymbol{\mathcal{J}} = -\left(\frac{\varepsilon_0 L^3}{2\omega^{(n)}}\right) \mathrm{Im}\Bigg\{&\left(\frac{k_y^{(m)}\hat{\boldsymbol{z}} - k_z^{(m)}\hat{\boldsymbol{y}}}{k_x^{(m)} - k_x^{(n)}}\right) \boldsymbol{E}_0^{(m)} \cdot \boldsymbol{E}_0^{(m_x \pm 1, m_y, m_z)*} + \left(\frac{k_z^{(m)}\hat{\boldsymbol{x}} - k_x^{(m)}\hat{\boldsymbol{z}}}{k_y^{(m)} - k_y^{(n)}}\right) \boldsymbol{E}_0^{(m)} \cdot \boldsymbol{E}_0^{(m_x, m_y \pm 1, m_z)*} \\ &+ \left(\frac{k_x^{(m)}\hat{\boldsymbol{y}} - k_y^{(m)}\hat{\boldsymbol{x}}}{k_z^{(m)} - k_z^{(n)}}\right) \boldsymbol{E}_0^{(m)} \cdot \boldsymbol{E}_0^{(m_x, m_y, m_z \pm 1)*} + \left(E_{0x}^{(m)} E_{0y}^{(m_x \pm 1, m_y, m_z)*}\hat{\boldsymbol{z}} - E_{0x}^{(m)} E_{0z}^{(m_x \pm 1, m_y, m_z)*}\hat{\boldsymbol{y}}\right) \\ &+ \left(E_{0y}^{(m)} E_{0z}^{(m_x, m_y \pm 1, m_z)*}\hat{\boldsymbol{x}} - E_{0y}^{(m)} E_{0x}^{(m_x, m_y \pm 1, m_z)*}\hat{\boldsymbol{z}}\right) + \left(E_{0z}^{(m)} E_{0x}^{(m_x, m_y, m_z \pm 1)*}\hat{\boldsymbol{y}} - E_{0z}^{(m)} E_{0y}^{(m_x, m_y, m_z \pm 1)*}\hat{\boldsymbol{x}}\right)\Bigg\}.\end{aligned} \tag{9}$$

It should be emphasized that Eq.(9) contains the contributions to the angular momentum $\boldsymbol{\mathcal{J}}$ by the $E$-field of the plane-wave $\boldsymbol{m} = (m_x, m_y, m_z)$ and the $B$-fields of the six adjacent plane-waves $\boldsymbol{n}_{1,2} = (m_x \pm 1, m_y, m_z)$, $\boldsymbol{n}_{3,4} = (m_x, m_y \pm 1, m_z)$, and $\boldsymbol{n}_{5,6} = (m_x, m_y, m_z \pm 1)$. Assuming the adjacent plane-wave amplitudes vary slowly with small changes in $\boldsymbol{k}$, the last three terms inside the curly brackets of Eq.(9) can be further simplified (upon averaging over the $\pm 1$ pairs) to yield

$$\left(E_{0y}^{(m)} E_{0z}^{(m)*} - E_{0z}^{(m)} E_{0y}^{(m)*}\right)\hat{\boldsymbol{x}} + \left(E_{0z}^{(m)} E_{0x}^{(m)*} - E_{0x}^{(m)} E_{0z}^{(m)*}\right)\hat{\boldsymbol{y}} + \left(E_{0x}^{(m)} E_{0y}^{(m)*} - E_{0y}^{(m)} E_{0x}^{(m)*}\right)\hat{\boldsymbol{z}}. \tag{10}$$

Substituting the above expression into Eq.(9) now yields the spin contribution $\boldsymbol{\mathcal{S}}$ to the total angular momentum $\boldsymbol{\mathcal{J}}$, as follows:

$$\boldsymbol{\mathcal{S}} \cong -\left(\frac{\varepsilon_0 L^3}{\omega^{(m)}}\right) \mathrm{Im}\left(\boldsymbol{E}_0^{(m)} \times \boldsymbol{E}_0^{(m)*}\right). \tag{11}$$

The assumption that adjacent plane-wave amplitudes vary slowly with changes in $\boldsymbol{k}$ also enables one to further simplify the first three terms inside the curly brackets of Eq.(9) to arrive at a streamlined expression for the orbital angular momentum $\boldsymbol{\mathcal{L}}$, namely,

$$\begin{aligned}\boldsymbol{\mathcal{L}} \cong \left(\frac{\varepsilon_0 L^3}{\omega^{(m)}}\right)\mathrm{Im}\big\{&\big(k_y^{(m)}\hat{\boldsymbol{z}} - k_z^{(m)}\hat{\boldsymbol{y}}\big)\big(E_{0x}^{(m)}\partial_{kx}E_{0x}^{(m)*} + E_{0y}^{(m)}\partial_{kx}E_{0y}^{(m)*} + E_{0z}^{(m)}\partial_{kx}E_{0z}^{(m)*}\big)\\
&+\big(k_z^{(m)}\hat{\boldsymbol{x}} - k_x^{(m)}\hat{\boldsymbol{z}}\big)\big(E_{0x}^{(m)}\partial_{ky}E_{0x}^{(m)*} + E_{0y}^{(m)}\partial_{ky}E_{0y}^{(m)*} + E_{0z}^{(m)}\partial_{ky}E_{0z}^{(m)*}\big)\\
&+\big(k_x^{(m)}\hat{\boldsymbol{y}} - k_y^{(m)}\hat{\boldsymbol{x}}\big)\big(E_{0x}^{(m)}\partial_{kz}E_{0x}^{(m)*} + E_{0y}^{(m)}\partial_{kz}E_{0y}^{(m)*} + E_{0z}^{(m)}\partial_{kz}E_{0z}^{(m)*}\big)\big\}\\
= -\left(\frac{\varepsilon_0 L^3}{\omega^{(m)}}\right)\mathrm{Im}\big\{&E_{0x}^{(m)}\big(k_y^{(m)}\partial_{kz}E_{0x}^{(m)*} - k_z^{(m)}\partial_{ky}E_{0x}^{(m)*}\big)\hat{\boldsymbol{x}} + E_{0y}^{(m)}\big(k_y^{(m)}\partial_{kz}E_{0y}^{(m)*} - k_z^{(m)}\partial_{ky}E_{0y}^{(m)*}\big)\hat{\boldsymbol{x}}\\
&+E_{0z}^{(m)}\big(k_y^{(m)}\partial_{kz}E_{0z}^{(m)*} - k_z^{(m)}\partial_{ky}E_{0z}^{(m)*}\big)\hat{\boldsymbol{x}}\big\}\\
&+E_{0x}^{(m)}\big(k_z^{(m)}\partial_{kx}E_{0x}^{(m)*} - k_x^{(m)}\partial_{kz}E_{0x}^{(m)*}\big)\hat{\boldsymbol{y}} + E_{0y}^{(m)}\big(k_z^{(m)}\partial_{kx}E_{0y}^{(m)*} - k_x^{(m)}\partial_{kz}E_{0y}^{(m)*}\big)\hat{\boldsymbol{y}}\\
&+E_{0z}^{(m)}\big(k_z^{(m)}\partial_{kx}E_{0z}^{(m)*} - k_x^{(m)}\partial_{kz}E_{0z}^{(m)*}\big)\hat{\boldsymbol{y}}\\
&+E_{0x}^{(m)}\big(k_x^{(m)}\partial_{ky}E_{0x}^{(m)*} - k_y^{(m)}\partial_{kx}E_{0x}^{(m)*}\big)\hat{\boldsymbol{z}} + E_{0y}^{(m)}\big(k_x^{(m)}\partial_{ky}E_{0y}^{(m)*} - k_y^{(m)}\partial_{kx}E_{0y}^{(m)*}\big)\hat{\boldsymbol{z}}\\
&+E_{0z}^{(m)}\big(k_x^{(m)}\partial_{ky}E_{0z}^{(m)*} - k_y^{(m)}\partial_{kx}E_{0z}^{(m)*}\big)\hat{\boldsymbol{z}}\big\}\\
= -\left(\frac{\varepsilon_0 L^3}{\omega^{(m)}}\right)\mathrm{Im}\big\{&E_{0x}^{(m)}\big(\boldsymbol{k}^{(m)}\times\boldsymbol{\nabla}_k E_{0x}^{(m)*}\big) + E_{0y}^{(m)}\big(\boldsymbol{k}^{(m)}\times\boldsymbol{\nabla}_k E_{0y}^{(m)*}\big) + E_{0z}^{(m)}\big(\boldsymbol{k}^{(m)}\times\boldsymbol{\nabla}_k E_{0z}^{(m)*}\big)\big\}. \qquad (12)\end{aligned}$$

An *incorrect* enhancement of both $\boldsymbol{\mathcal{S}}$ and $\boldsymbol{\mathcal{L}}$ by a factor of 2 notwithstanding, Eqs. (11) and (12) are in general agreement with standard results obtained for classical EM fields possessing a continuous and differentiable spatio-temporal spectrum in the $(\boldsymbol{k},\omega)$ domain.[1] It is unfortunate that the final results of this fairly simple calculation of optical angular momenta turn out to be twice as large as their correct values. The appearance of this extraneous factor of 2, arising from the necessity of accounting for six (rather than three)[§] nearest-neighbors of each plane-wave in the discretized $k$-space in accordance with Eq.(9), is unfortunate but appears to be unavoidable.

As for the energy content of the fields, the dot-products $\boldsymbol{E}^{(m)}\cdot\boldsymbol{E}^{(n)}$ and $\boldsymbol{B}^{(m)}\cdot\boldsymbol{B}^{(n)}$ both integrate to zero over the $L^3$ volume of the $xyz$ space if $\boldsymbol{m}\neq\boldsymbol{n}$, in which case only individual plane-waves (as opposed to pairs) contribute to the $\boldsymbol{E}$ and $\boldsymbol{B}$ field energies, as follows:

$$\begin{aligned}\mathcal{E}_{\text{electric}}^{(n)} &= \tfrac{1}{2}\varepsilon_0\iiint_{-L/2}^{L/2}\boldsymbol{E}^{(n)}(\boldsymbol{r},t)\cdot\boldsymbol{E}^{(n)}(\boldsymbol{r},t)\mathrm{d}x\mathrm{d}y\mathrm{d}z = \tfrac{1}{2}\varepsilon_0\iiint_{-L/2}^{L/2}\tfrac{1}{2}\mathrm{Re}\big\{\boldsymbol{E}_0^{(n)}\cdot\boldsymbol{E}_0^{(n)*}\big\}\mathrm{d}x\mathrm{d}y\mathrm{d}z\\
&= \tfrac{1}{4}\varepsilon_0 L^3\boldsymbol{E}_0^{(n)}\cdot\boldsymbol{E}_0^{(n)*}. \qquad (13)\end{aligned}$$

$$\begin{aligned}\mathcal{E}_{\text{magnetic}}^{(n)} &= \tfrac{1}{2}\mu_0^{-1}\iiint_{-L/2}^{L/2}\boldsymbol{B}^{(n)}(\boldsymbol{r},t)\cdot\boldsymbol{B}^{(n)}(\boldsymbol{r},t)\mathrm{d}x\mathrm{d}y\mathrm{d}z = \tfrac{1}{2}\mu_0^{-1}\iiint_{-L/2}^{L/2}\tfrac{1}{2}\mathrm{Re}\big\{\boldsymbol{B}_0^{(n)}\cdot\boldsymbol{B}_0^{(n)*}\big\}\mathrm{d}x\mathrm{d}y\mathrm{d}z\\
&= \tfrac{1}{4}\mu_0^{-1}L^3\mathrm{Re}\big\{\boldsymbol{B}_0^{(n)}\cdot\boldsymbol{B}_0^{(n)*}\big\} = \frac{L^3}{4\mu_0\omega^{(n)2}}\big(\boldsymbol{k}^{(n)}\times\boldsymbol{E}_0^{(n)}\big)\cdot\big(\boldsymbol{k}^{(n)}\times\boldsymbol{E}_0^{(n)*}\big)\\
&= \frac{L^3}{4\mu_0\omega^{(n)2}}\Big[\big(\boldsymbol{k}^{(n)}\cdot\boldsymbol{k}^{(n)}\big)\big(\boldsymbol{E}_0^{(n)}\cdot\boldsymbol{E}_0^{(n)*}\big) - \overset{0}{\big(\boldsymbol{k}^{(n)}\cdot\boldsymbol{E}_0^{(n)*}\big)}\overset{0}{\big(\boldsymbol{k}^{(n)}\cdot\boldsymbol{E}_0^{(n)}\big)}\Big] = \tfrac{1}{4}\varepsilon_0 L^3\boldsymbol{E}_0^{(n)}\cdot\boldsymbol{E}_0^{(n)*}. \qquad (14)\end{aligned}$$

The total (i.e., electric plus magnetic) energy content of each plane-wave is thus seen to be $\mathcal{E}^{(n)} = \tfrac{1}{2}\varepsilon_0 L^3\boldsymbol{E}_0^{(n)}\cdot\boldsymbol{E}_0^{(n)*}$.

---

[§] The six nearest-neighbors of the plane-wave with index triplet $(m_x, m_y, m_z)$ appear in Eq.(9) with their index triplets specified as $(m_x\pm 1, m_y\pm 1, m_z\pm 1)$. The extraneous factor of 2 in the final expression of $\boldsymbol{\mathcal{J}}$ is caused by optical interference between the plane-wave $\boldsymbol{m}$ and *both* its adjacent neighbors along each direction. The offending factor of 2 would disappear if the plane-wave $\boldsymbol{m}$ interfered with only one neighbor in each direction.

For right- and left-circularly-polarized (RCP and LCP) plane-waves, the $E$-field amplitude is $E_0\hat{\boldsymbol{e}} = E_0(\boldsymbol{e}' \pm \mathrm{i}\boldsymbol{e}'')$, with the plus sign for RCP and the minus sign for LCP, where $\boldsymbol{e}' \cdot \boldsymbol{e}'' = 0$ and $\hat{\boldsymbol{e}} \cdot \hat{\boldsymbol{e}}^* = \boldsymbol{e}' \cdot \boldsymbol{e}' + \boldsymbol{e}'' \cdot \boldsymbol{e}'' = ½ + ½ = 1$. The complex unit-vector $\hat{\boldsymbol{e}}$ is also orthogonal to the corresponding $k$-vector; that is, $\hat{\boldsymbol{\kappa}} \cdot \hat{\boldsymbol{e}} = 0$. While the total energy content of each such plane-wave is $\mathcal{E} = ½\varepsilon_0 L^3|E_0|^2$, the spin angular momenta of the RCP and LCP plane-waves are found from Eq.(11) to be $\boldsymbol{\mathcal{S}} = \pm(\varepsilon_0 L^3|E_0|^2/\omega)\hat{\boldsymbol{\kappa}}$. Thus, the ratio of energy to spin angular momentum for these circularly-polarized plane-waves is seen to be $\omega/2$ (instead of $\omega$), which is a consequence of the aforementioned fact that our expressions of angular momenta in Eqs.(9), (11) and (12) are incorrectly enhanced by a factor of 2.

**3. Interfering plane-waves inside a large sphere**. Suspecting that discretization in Cartesian coordinates might be responsible for the erroneous results of Sec.2, we switch to spherical coordinates and examine a pair of propagating plane-waves in free space with $k$-vectors $\boldsymbol{k}_1$ and $\boldsymbol{k}_2$, but with the same frequency $\omega$ (i.e., $|\boldsymbol{k}_1| = |\boldsymbol{k}_2| = \omega/c$). The relevant integral for computing the $\boldsymbol{E}$ and $\boldsymbol{B}$ field energies over a spherical volume of radius $R$ is seen to be

$$\iiint e^{\mathrm{i}(\boldsymbol{k}_1-\boldsymbol{k}_2)\cdot\boldsymbol{r}}\mathrm{d}\boldsymbol{r} = \int_{r=0}^{R}\int_{\varphi=0}^{\pi} 2\pi r^2 \sin\varphi\, e^{\mathrm{i}|\boldsymbol{k}_1-\boldsymbol{k}_2|r\cos\varphi}\mathrm{d}\varphi\mathrm{d}r$$

$\hat{\boldsymbol{\kappa}} = (\boldsymbol{k}_1 - \boldsymbol{k}_2)/|\boldsymbol{k}_1 - \boldsymbol{k}_2|$ is the axis of symmetry; see Fig.1.

d$\boldsymbol{r}$ stands for d$x$d$y$d$z$

$$= \frac{4\pi}{|\boldsymbol{k}_1-\boldsymbol{k}_2|}\int_{r=0}^{R} r\sin(|\boldsymbol{k}_1-\boldsymbol{k}_2|r)\,\mathrm{d}r = \frac{4\pi}{|\boldsymbol{k}_1-\boldsymbol{k}_2|^3}\int_{x=0}^{|\boldsymbol{k}_1-\boldsymbol{k}_2|R} x\sin x\,\mathrm{d}x. \tag{15}$$

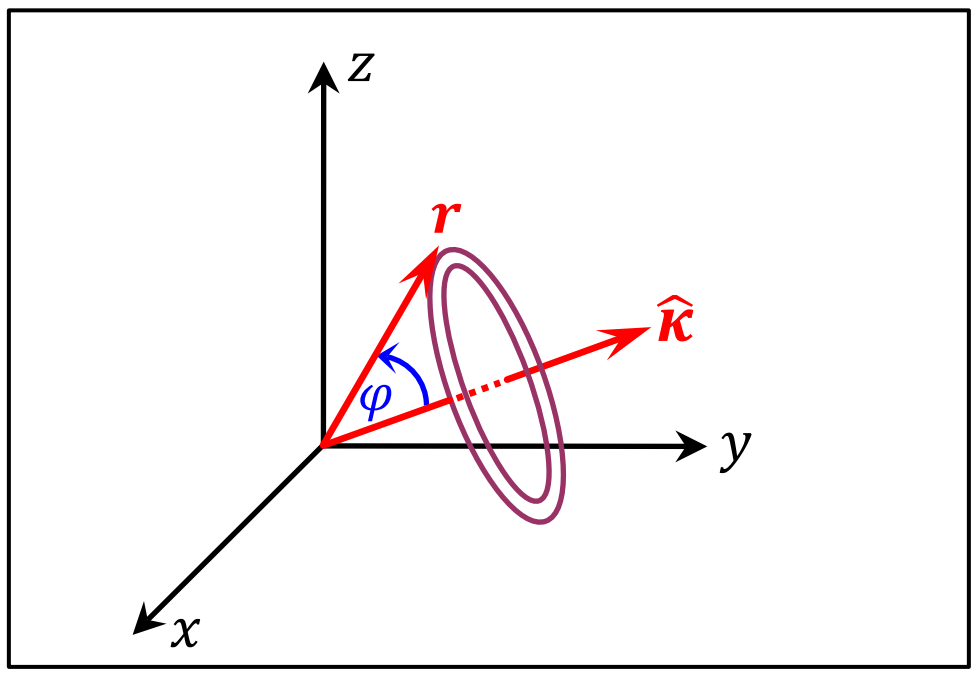

**Fig.1**. In the $xyz$ coordinate system, integrals over a spherical volume of radius $R$ are carried out first around a ring of radius $r\sin\varphi$ and thickness $r\mathrm{d}\varphi\mathrm{d}r$ centered on the axis of symmetry $\hat{\boldsymbol{\kappa}}$, then over the angle $\varphi$ from 0 to $\pi$, and finally over $r$ from 0 to $R$.

The elementary integral on the right-hand side of Eq.(15) is readily evaluated, as follows:

$$\int_{x=0}^{|\boldsymbol{k}_1-\boldsymbol{k}_2|R} x\sin(x)\,\mathrm{d}x = -x\cos x\Big|_0^{|\boldsymbol{k}_1-\boldsymbol{k}_2|R} + \int_{x=0}^{|\boldsymbol{k}_1-\boldsymbol{k}_2|R}\cos(x)\,\mathrm{d}x$$

$$= \sin(|\boldsymbol{k}_1-\boldsymbol{k}_2|R) - |\boldsymbol{k}_1-\boldsymbol{k}_2|R\cos(|\boldsymbol{k}_1-\boldsymbol{k}_2|R). \tag{16}$$

In the limit when $|\boldsymbol{k}_1 - \boldsymbol{k}_2|R \to 0$, the above expression approaches $(|\boldsymbol{k}_1 - \boldsymbol{k}_2|R)^3/3$ and, therefore, $\iiint e^{\mathrm{i}(\boldsymbol{k}_1-\boldsymbol{k}_2)\cdot\boldsymbol{r}}\mathrm{d}\boldsymbol{r} \to 4\pi R^3/3$, namely, the volume $V$ of the sphere. Similarly, the relevant integral for computing the angular momentum of the plane-wave-pair over the spherical volume is

$$\iiint \boldsymbol{r}e^{\mathrm{i}(\boldsymbol{k}_1-\boldsymbol{k}_2)\cdot\boldsymbol{r}}\mathrm{d}\boldsymbol{r} = \frac{\boldsymbol{k}_1-\boldsymbol{k}_2}{|\boldsymbol{k}_1-\boldsymbol{k}_2|}\int_{r=0}^{R}\int_{\varphi=0}^{\pi} 2\pi r^3\sin\varphi\cos\varphi\, e^{\mathrm{i}|\boldsymbol{k}_1-\boldsymbol{k}_2|r\cos\varphi}\mathrm{d}\varphi\mathrm{d}r$$

see Fig.1

$$= \frac{\mathrm{i}4\pi(\boldsymbol{k}_1-\boldsymbol{k}_2)}{|\boldsymbol{k}_1-\boldsymbol{k}_2|^3}\int_{r=0}^{R} r[\sin(|\boldsymbol{k}_1-\boldsymbol{k}_2|r) - |\boldsymbol{k}_1-\boldsymbol{k}_2|r\cos(|\boldsymbol{k}_1-\boldsymbol{k}_2|r)]\mathrm{d}r$$

$$= \frac{\mathrm{i}4\pi(\boldsymbol{k}_1-\boldsymbol{k}_2)}{|\boldsymbol{k}_1-\boldsymbol{k}_2|^5}\int_{x=0}^{|\boldsymbol{k}_1-\boldsymbol{k}_2|R} x(\sin x - x\cos x)\mathrm{d}x. \tag{17}$$

Evaluating the elementary integral on the right-hand side of Eq.(17), we find

$$\int_{x=0}^{|\boldsymbol{k}_1-\boldsymbol{k}_2|R} x^2\cos x\,\mathrm{d}x = x^2\sin x\big|_0^{|\boldsymbol{k}_1-\boldsymbol{k}_2|R} - \int_{x=0}^{|\boldsymbol{k}_1-\boldsymbol{k}_2|R} 2x\sin x\,\mathrm{d}x. \tag{18}$$

$$\int_{x=0}^{|\boldsymbol{k}_1-\boldsymbol{k}_2|R} x(\sin x - x\cos x)\mathrm{d}x = 3\int_{x=0}^{|\boldsymbol{k}_1-\boldsymbol{k}_2|R} x\sin x\,\mathrm{d}x - x^2\sin x\big|_0^{|\boldsymbol{k}_1-\boldsymbol{k}_2|R}$$
$$= 3\sin(|\boldsymbol{k}_1-\boldsymbol{k}_2|R) - 3|\boldsymbol{k}_1-\boldsymbol{k}_2|R\cos(|\boldsymbol{k}_1-\boldsymbol{k}_2|R) - (|\boldsymbol{k}_1-\boldsymbol{k}_2|R)^2\sin(|\boldsymbol{k}_1-\boldsymbol{k}_2|R). \tag{19}$$

In the limit when $|\boldsymbol{k}_1-\boldsymbol{k}_2|R \to 0$, the above expression approaches $(|\boldsymbol{k}_1-\boldsymbol{k}_2|R)^5/15$ and, therefore,

$$\iiint \boldsymbol{r}e^{\mathrm{i}(\boldsymbol{k}_1-\boldsymbol{k}_2)\cdot\boldsymbol{r}}\mathrm{d}\boldsymbol{r} \to \mathrm{i}4\pi R^5(\boldsymbol{k}_1-\boldsymbol{k}_2)/15 = \mathrm{i}VR^2(\boldsymbol{k}_1-\boldsymbol{k}_2)/5. \tag{20}$$

Here, $V = 4\pi R^3/3$ is the sphere's volume. A reasonable estimate for $|\boldsymbol{k}_1-\boldsymbol{k}_2|$ that comes close to the desired limit $|\boldsymbol{k}_1-\boldsymbol{k}_2|R \to 0$ is $|\boldsymbol{k}_1-\boldsymbol{k}_2| = \alpha/R$, with $0 < \alpha \lesssim 0.1$. Introducing the unit-vector $\hat{\boldsymbol{\kappa}} = (\boldsymbol{k}_1-\boldsymbol{k}_2)/|\boldsymbol{k}_1-\boldsymbol{k}_2|$, we may now write: $\iiint \boldsymbol{r}e^{\mathrm{i}(\boldsymbol{k}_1-\boldsymbol{k}_2)\cdot\boldsymbol{r}}\mathrm{d}\boldsymbol{r} \to \mathrm{i}\alpha^2V\hat{\boldsymbol{\kappa}}/(5|\boldsymbol{k}_1-\boldsymbol{k}_2|)$.

Next, we compute the $E$-field energy of the plane-wave pair $\boldsymbol{E}_1e^{\mathrm{i}(\boldsymbol{k}_1\cdot\boldsymbol{r}-\omega t)}$, $\boldsymbol{E}_2e^{\mathrm{i}(\boldsymbol{k}_2\cdot\boldsymbol{r}-\omega t)}$ by integrating the pair's combined energy-density over the volume of the sphere of radius $R$; that is,

$$\mathcal{E}_{\text{electric}} = \iiint \tfrac{1}{4}\varepsilon_0\mathrm{Re}\{(\boldsymbol{E}_1e^{\mathrm{i}\boldsymbol{k}_1\cdot\boldsymbol{r}} + \boldsymbol{E}_2e^{\mathrm{i}\boldsymbol{k}_2\cdot\boldsymbol{r}})\cdot(\boldsymbol{E}_1^*e^{-\mathrm{i}\boldsymbol{k}_1\cdot\boldsymbol{r}} + \boldsymbol{E}_2^*e^{-\mathrm{i}\boldsymbol{k}_2\cdot\boldsymbol{r}})\}\mathrm{d}\boldsymbol{r}$$
$$= \tfrac{1}{3}\pi\varepsilon_0R^3(\boldsymbol{E}_1\cdot\boldsymbol{E}_1^* + \boldsymbol{E}_2\cdot\boldsymbol{E}_2^*) + \tfrac{1}{4}\varepsilon_0\iiint[\boldsymbol{E}_1\cdot\boldsymbol{E}_2^*e^{\mathrm{i}(\boldsymbol{k}_1-\boldsymbol{k}_2)\cdot\boldsymbol{r}} + \boldsymbol{E}_2\cdot\boldsymbol{E}_1^*e^{-\mathrm{i}(\boldsymbol{k}_1-\boldsymbol{k}_2)\cdot\boldsymbol{r}}]\mathrm{d}\boldsymbol{r}. \tag{21}$$

In the limit when $|\boldsymbol{k}_1-\boldsymbol{k}_2|R \to 0$, both integrals (over the spherical volume) of the complex exponential functions $e^{\pm\mathrm{i}(\boldsymbol{k}_1-\boldsymbol{k}_2)\cdot\boldsymbol{r}}$ approach $V = 4\pi R^3/3$, yielding

$$\mathcal{E}_{\text{electric}} \to \tfrac{1}{3}\pi\varepsilon_0R^3(\boldsymbol{E}_1\cdot\boldsymbol{E}_1^* + \boldsymbol{E}_2\cdot\boldsymbol{E}_2^* + \boldsymbol{E}_1\cdot\boldsymbol{E}_2^* + \boldsymbol{E}_2\cdot\boldsymbol{E}_1^*) = \tfrac{1}{4}\varepsilon_0V(\boldsymbol{E}_1+\boldsymbol{E}_2)\cdot(\boldsymbol{E}_1+\boldsymbol{E}_2)^*. \tag{22}$$

It is seen that, in the limit when $|\boldsymbol{k}_1-\boldsymbol{k}_2|R \to 0$, the two plane-waves tend to overlap each other, behaving as a single, monochromatic plane-wave. However, before reaching this limit, Eq.(16) governs the sinusoidal variations of the $E$-field energy contained in the sphere as a function of its radius $R$. It is unfortunate that, in cases corresponding to $|\boldsymbol{k}_1-\boldsymbol{k}_2| = \alpha/R$ (with $\alpha \lesssim 0.1$), the energy content of the sphere does *not* equal the sum of the energies of the individual plane-waves, since this is the limit in which the plane-wave-pair's angular momentum reaches a definite stable value independently of $R$ (see below). In hindsight, however, the limit value of $\mathcal{E}_{\text{electric}}$ as given by Eq.(22) should not come as a surprise, considering that, with $\alpha \lesssim 0.1$, the sphere of radius $R$ contains only a small fraction of a single interference fringe.

As for the angular momentum $\boldsymbol{\mathcal{J}}$ of the pair of plane-waves, we begin by integrating the angular-momentum-density $\boldsymbol{j}(\boldsymbol{r},t) = \boldsymbol{r}\times\varepsilon_0\mathrm{Re}[\boldsymbol{E}(\boldsymbol{r},t)]\times\mathrm{Re}[\boldsymbol{B}(\boldsymbol{r},t)]$ over the spherical volume of radius $R$. Dropping the high-frequency temporal oscillations (frequency $= 2\omega$), we arrive at

$$\boldsymbol{\mathcal{J}} = \iiint \boldsymbol{r}\times\tfrac{1}{2}\varepsilon_0\mathrm{Re}[(\boldsymbol{E}_1e^{\mathrm{i}\boldsymbol{k}_1\cdot\boldsymbol{r}} + \boldsymbol{E}_2e^{\mathrm{i}\boldsymbol{k}_2\cdot\boldsymbol{r}})\times(\boldsymbol{B}_1^*e^{-\mathrm{i}\boldsymbol{k}_1\cdot\boldsymbol{r}} + \boldsymbol{B}_2^*e^{-\mathrm{i}\boldsymbol{k}_2\cdot\boldsymbol{r}})]\mathrm{d}\boldsymbol{r}$$
$$= \tfrac{1}{2}\varepsilon_0\mathrm{Re}\iiint \boldsymbol{r}\times[\boldsymbol{E}_1\times\boldsymbol{B}_2^*e^{\mathrm{i}(\boldsymbol{k}_1-\boldsymbol{k}_2)\cdot\boldsymbol{r}} + \boldsymbol{E}_2\times\boldsymbol{B}_1^*e^{-\mathrm{i}(\boldsymbol{k}_1-\boldsymbol{k}_2)\cdot\boldsymbol{r}}]\mathrm{d}\boldsymbol{r} \quad \leftarrow \text{see Eq.(17)}$$
$$= -\frac{2\pi\varepsilon_0(\boldsymbol{k}_1-\boldsymbol{k}_2)}{|\boldsymbol{k}_1-\boldsymbol{k}_2|^5}\times\mathrm{Im}(\boldsymbol{E}_1\times\boldsymbol{B}_2^* - \boldsymbol{E}_2\times\boldsymbol{B}_1^*)\int_{x=0}^{|\boldsymbol{k}_1-\boldsymbol{k}_2|R} x(\sin x - x\cos x)\mathrm{d}x. \tag{23}$$

The remaining integral that appears on the right-hand side of Eq.(23) is given by Eq.(19). The interminable oscillations of this integral with an increasing radius $R$ cannot be suppressed unless we choose to work in the limit where $|\boldsymbol{k}_1-\boldsymbol{k}_2|R \to 0$ or, equivalently, set $|\boldsymbol{k}_1-\boldsymbol{k}_2| = \alpha/R$ with, say, $\alpha \lesssim 0.1$. Invoking the limiting value of the integral as given by Eq.(20), and recalling that the unit-vector $\hat{\boldsymbol{\kappa}}$ is the normalized $\boldsymbol{k}_1 - \boldsymbol{k}_2$, we find (for sufficiently small $\alpha$) that the stable value of the angular momentum $\boldsymbol{\mathcal{J}}$ is given by

$$\boldsymbol{\mathcal{J}} \cong \varepsilon_0 V(R^2/10)(\boldsymbol{k}_1 - \boldsymbol{k}_2) \times \text{Im}(\boldsymbol{E}_2 \times \boldsymbol{B}_1^* - \boldsymbol{E}_1 \times \boldsymbol{B}_2^*) \leftarrow \boxed{|\boldsymbol{k}_1 - \boldsymbol{k}_2| = \alpha/R}$$

$$= (\varepsilon_0 V\alpha^2/10\omega)\hat{\boldsymbol{\kappa}} \times \text{Im}[\boldsymbol{E}_2 \times (\boldsymbol{k}_1 \times \boldsymbol{E}_1^*) - \boldsymbol{E}_1 \times (\boldsymbol{k}_2 \times \boldsymbol{E}_2^*)]/|\boldsymbol{k}_1 - \boldsymbol{k}_2|$$

$$= (\varepsilon_0 V\alpha^2/10\omega)\hat{\boldsymbol{\kappa}} \times \text{Im}[(\boldsymbol{E}_2 \cdot \boldsymbol{E}_1^*)\boldsymbol{k}_1 - (\boldsymbol{E}_2 \cdot \boldsymbol{k}_1)\boldsymbol{E}_1^* - (\boldsymbol{E}_1 \cdot \boldsymbol{E}_2^*)\boldsymbol{k}_2 + (\boldsymbol{E}_1 \cdot \boldsymbol{k}_2)\boldsymbol{E}_2^*]/|\boldsymbol{k}_1 - \boldsymbol{k}_2|. \quad (24)$$

Given that $\boldsymbol{k}_1 \cdot \boldsymbol{E}_1 = 0$ and $\boldsymbol{k}_2 \cdot \boldsymbol{E}_2 = 0$ (according to Maxwell's 1st equation, $\boldsymbol{\nabla} \cdot \boldsymbol{E}(\boldsymbol{r}, t) = 0$), one may replace $\boldsymbol{E}_1 \cdot \boldsymbol{k}_2$ with $\boldsymbol{E}_1 \cdot (\boldsymbol{k}_2 - \boldsymbol{k}_1)$ and, similarly, $\boldsymbol{E}_2 \cdot \boldsymbol{k}_1$ with $\boldsymbol{E}_2 \cdot (\boldsymbol{k}_1 - \boldsymbol{k}_2)$. It is also permissible to replace $\boldsymbol{E}_2 \cdot \boldsymbol{E}_1^*$ with $(\boldsymbol{E}_2 - \boldsymbol{E}_1) \cdot \boldsymbol{E}_1^*$ and $\boldsymbol{E}_1 \cdot \boldsymbol{E}_2^*$ with $(\boldsymbol{E}_1 - \boldsymbol{E}_2) \cdot \boldsymbol{E}_2^*$, ostensibly because the added terms $\boldsymbol{E}_1 \cdot \boldsymbol{E}_1^*$ and $\boldsymbol{E}_2 \cdot \boldsymbol{E}_2^*$ are real and, therefore, do not alter the imaginary part of the bracketed expression in Eq.(24). We thus have

$$\boldsymbol{\mathcal{J}} \cong -\left(\frac{\varepsilon_0 V\alpha^2}{10\omega}\right) \text{Im}\left\{\frac{(\boldsymbol{E}_1 - \boldsymbol{E}_2)\cdot \boldsymbol{E}_1^*}{|\boldsymbol{k}_1 - \boldsymbol{k}_2|}(\hat{\boldsymbol{\kappa}} \times \boldsymbol{k}_1) + \frac{(\boldsymbol{E}_1 - \boldsymbol{E}_2)\cdot \boldsymbol{E}_2^*}{|\boldsymbol{k}_1 - \boldsymbol{k}_2|}(\hat{\boldsymbol{\kappa}} \times \boldsymbol{k}_2) + (\boldsymbol{E}_1 \cdot \hat{\boldsymbol{\kappa}})(\hat{\boldsymbol{\kappa}} \times \boldsymbol{E}_2^*) + (\boldsymbol{E}_2 \cdot \hat{\boldsymbol{\kappa}})(\hat{\boldsymbol{\kappa}} \times \boldsymbol{E}_1^*)\right\}. \quad (25)$$

$\uparrow \boxed{|\boldsymbol{k}_1 - \boldsymbol{k}_2| = \alpha/R}$

In the continuum limit, where the $E$-field varies slowly with $\boldsymbol{k}$, the first two terms inside the curly brackets of Eq.(25) approach $(\boldsymbol{E}_1^* \cdot \partial_{\hat{\boldsymbol{\kappa}}}\boldsymbol{E}_1)\hat{\boldsymbol{\kappa}} \times \boldsymbol{k}_1 + (\boldsymbol{E}_2^* \cdot \partial_{\hat{\boldsymbol{\kappa}}}\boldsymbol{E}_2)\hat{\boldsymbol{\kappa}} \times \boldsymbol{k}_2$. Aside from the leading coefficient $\alpha^2/10$ appearing in Eq.(25), this expression of the orbital angular momentum $\boldsymbol{\mathcal{L}}$ for the pair of plane-waves under consideration agrees with the general result obtained in Ref.[3].

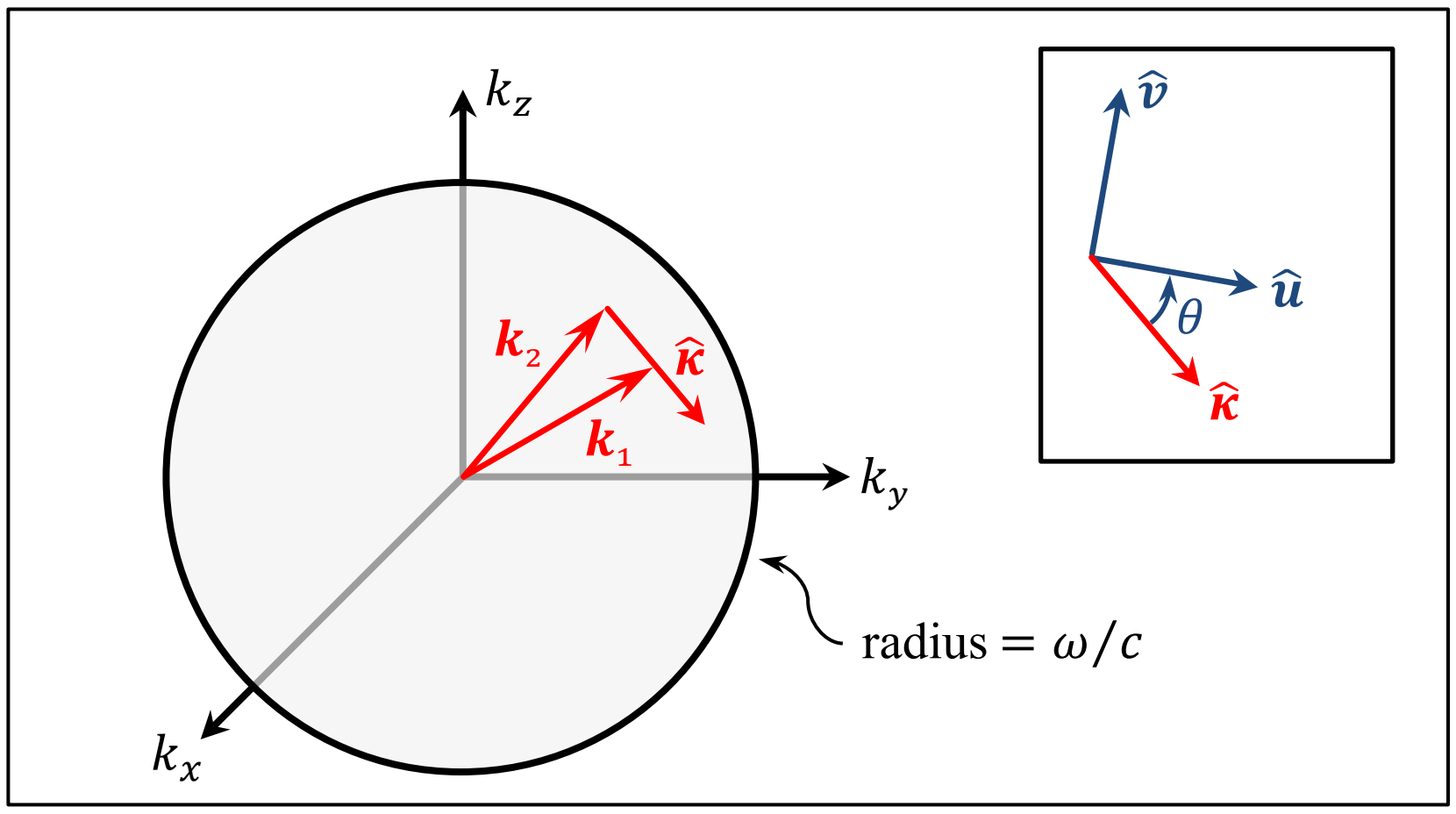

**Fig.2**. The wave-vectors $\boldsymbol{k}_1$ and $\boldsymbol{k}_2$ represent two nearby points in the $k$-space on the surface of a sphere of radius $|\boldsymbol{k}| = \omega/c$. The unit-vector connecting the two points on the spherical surface is $\hat{\boldsymbol{\kappa}} = (\boldsymbol{k}_1 - \boldsymbol{k}_2)/|\boldsymbol{k}_1 - \boldsymbol{k}_2|$. At any given point on the sphere's surface, the surface normal $\hat{\boldsymbol{n}}$ (not shown) is the unit-vector along the direction of the local $k$-vector. The $E$-field of each plane-wave is given by $\boldsymbol{E}(\boldsymbol{r}, t) = (E'\hat{\boldsymbol{u}} + E''\hat{\boldsymbol{v}})\exp[\text{i}(\boldsymbol{k} \cdot \boldsymbol{r} - \omega t)]$, where $E'$ and $E''$ are, in general, complex, and $(\hat{\boldsymbol{u}}, \hat{\boldsymbol{v}})$ is a pair of orthogonal unit-vectors perpendicular to the corresponding $k$-vector. The inset shows the relative orientations of $\hat{\boldsymbol{\kappa}}$, $\hat{\boldsymbol{u}}$, and $\hat{\boldsymbol{v}}$ within a plane perpendicular to the local $k$-vector.

The last two terms in Eq.(25) represent the spin angular momentum $\boldsymbol{\mathcal{S}}$ of the pair of plane-waves. To understand the characteristic feature of these terms, consider Fig.2, which shows $\boldsymbol{k}_1$ and $\boldsymbol{k}_2$ as two points on the surface of the $k$-space sphere of radius $|\boldsymbol{k}| = \omega/c$, along with the normalized vector $\boldsymbol{k}_1 - \boldsymbol{k}_2$, depicted as the unit-vector $\hat{\boldsymbol{\kappa}}$. The $E$-field of each plane-wave can be written as $\boldsymbol{E}(\boldsymbol{r},t) = (E'\hat{\boldsymbol{u}} + E''\hat{\boldsymbol{v}})\exp[\mathrm{i}(\boldsymbol{k}\cdot\boldsymbol{r} - \omega t)]$, where $(\hat{\boldsymbol{u}},\hat{\boldsymbol{v}})$ is a pair of mutually orthogonal unit-vectors, both being perpendicular to the plane-wave's $k$-vector. The $E$-field amplitudes $E'$ and $E''$ are, in general, complex-valued. The three unit-vectors $\hat{\boldsymbol{\kappa}}$, $\hat{\boldsymbol{u}}$, and $\hat{\boldsymbol{v}}$ become co-planar when $\boldsymbol{k}_1$ and $\boldsymbol{k}_2$ get to be sufficiently close to each other.

Returning to the third term $(\boldsymbol{E}_1\cdot\hat{\boldsymbol{\kappa}})\hat{\boldsymbol{\kappa}}\times\boldsymbol{E}_2^*$ inside the curly brackets of Eq.(25), we note that reversing the direction of $\hat{\boldsymbol{\kappa}}$ does not alter the value of this term, which indicates that $\boldsymbol{k}_2$ could be on either side of $\boldsymbol{k}_1$ (along the $\hat{\boldsymbol{\kappa}}$ direction). Thus, in the continuum limit, $\boldsymbol{E}_2^*$ approaches the average value of $\boldsymbol{E}^*$ between the two adjacent neighbors of $\boldsymbol{E}_1^*$, which is the same as $\boldsymbol{E}_1^*$. In other words, it is permissible (in the continuum limit) to equate $(\boldsymbol{E}_1\cdot\hat{\boldsymbol{\kappa}})\hat{\boldsymbol{\kappa}}\times\boldsymbol{E}_2^*$ with $(\boldsymbol{E}_1\cdot\hat{\boldsymbol{\kappa}})\hat{\boldsymbol{\kappa}}\times\boldsymbol{E}_1^*$. By the same token, the fourth term inside the curly brackets of Eq.(25) approaches $(\boldsymbol{E}_2\cdot\hat{\boldsymbol{\kappa}})(\hat{\boldsymbol{\kappa}}\times\boldsymbol{E}_2^*)$ in the continuum limit. We may now write

$$\begin{aligned}\mathrm{Im}[(\boldsymbol{E}_1\cdot\hat{\boldsymbol{\kappa}})\hat{\boldsymbol{\kappa}}\times\boldsymbol{E}_2^*] &\cong \mathrm{Im}\{[(E_1'\hat{\boldsymbol{u}} + E_1''\hat{\boldsymbol{v}})\cdot\hat{\boldsymbol{\kappa}}][\hat{\boldsymbol{\kappa}}\times(E_1'^*\hat{\boldsymbol{u}} + E_1''^*\hat{\boldsymbol{v}})]\}\\ &= \mathrm{Im}\{(E_1'\cos\theta - E_1''\sin\theta)(E_1'^*\sin\theta + E_1''^*\cos\theta)\hat{\boldsymbol{n}}\}\\ &= \mathrm{Im}(E_1'E_1''^*\cos^2\theta - E_1''E_1'^*\sin^2\theta)\hat{\boldsymbol{n}}\\ &= (\cos^2\theta + \sin^2\theta)\mathrm{Im}(E_1'E_1''^*)\hat{\boldsymbol{n}} = \tfrac{1}{2}\mathrm{Im}(\boldsymbol{E}_1\times\boldsymbol{E}_1^*). \qquad (26)\end{aligned}$$

A similar argument applies to the fourth term inside the curly brackets of Eq.(25). The overall angular momentum of the plane-wave pair thus becomes

$$\boldsymbol{\mathcal{J}} \cong \left(\frac{\varepsilon_0 V\alpha^2}{10\omega}\right)\mathrm{Im}\{(\boldsymbol{E}_1\cdot\partial_{\hat{\boldsymbol{\kappa}}}\boldsymbol{E}_1^*)\hat{\boldsymbol{\kappa}}\times\boldsymbol{k}_1 + (\boldsymbol{E}_2\cdot\partial_{\hat{\boldsymbol{\kappa}}}\boldsymbol{E}_2^*)\hat{\boldsymbol{\kappa}}\times\boldsymbol{k}_2 - \tfrac{1}{2}\boldsymbol{E}_1\times\boldsymbol{E}_1^* - \tfrac{1}{2}\boldsymbol{E}_2\times\boldsymbol{E}_2^*\}. \qquad (27)$$

If a third neighboring plane-wave $(\boldsymbol{k}_3,\omega)$ is brought in to similarly interact with the first plane-wave $(\boldsymbol{k}_1,\omega)$ in such a way that $(\boldsymbol{k}_1 - \boldsymbol{k}_3)\perp(\boldsymbol{k}_1 - \boldsymbol{k}_2)$, then the spin and orbital angular momenta contributed by the first plane-wave (in consequence of its interactions with the 2nd and 3rd plane-waves) will be nearly identical to those found in Ref.[3]. The only point of contention, however, will be the leading coefficient $\alpha^2/10$ in Eq.(27), which has no counterpart in Ref.[3].

**4. Integrating over the volume occupied by a single fringe**. Consider the pair of plane-waves $(\boldsymbol{E}_1,\boldsymbol{k}_1,\omega)$ and $(\boldsymbol{E}_2,\boldsymbol{k}_2,\omega)$, where $\boldsymbol{k}_{1,2} = \pm k_x\hat{\boldsymbol{x}} + k_z\hat{\boldsymbol{z}}$. The planar interference fringes formed by this pair are parallel to the $yz$-plane, with periodicity of $\pi/k_x$ along the $x$-axis, as depicted in Fig.3. Within a cubic volume $(L_x,L_y,L_z) = (\pi/k_x, 1.0, 1.0)$ centered at the origin, the $E$-field energy of a single fringe is found to be

$$\begin{aligned}\mathcal{E}_{\text{electric}} &= \tfrac{1}{4}\varepsilon_0\mathrm{Re}\int_{x=-\pi/2k_x}^{\pi/2k_x}[\boldsymbol{E}_1 e^{\mathrm{i}(k_x x + k_z z)} + \boldsymbol{E}_2 e^{\mathrm{i}(-k_x x + k_z z)}]\cdot[\boldsymbol{E}_1^* e^{-\mathrm{i}(k_x x + k_z z)} + \boldsymbol{E}_2^* e^{-\mathrm{i}(-k_x x + k_z z)}]\mathrm{d}x\\ &= (\pi\varepsilon_0/4k_x)(\boldsymbol{E}_1\cdot\boldsymbol{E}_1^* + \boldsymbol{E}_2\cdot\boldsymbol{E}_2^*) + \tfrac{1}{2}\varepsilon_0\mathrm{Re}[(\boldsymbol{E}_1\cdot\boldsymbol{E}_2^*)\underbrace{\int_{x=-\pi/2k_x}^{\pi/2k_x} e^{2\mathrm{i}k_x x}}_{0}\mathrm{d}x]\\ &= (\pi\varepsilon_0/4k_x)(\boldsymbol{E}_1\cdot\boldsymbol{E}_1^* + \boldsymbol{E}_2\cdot\boldsymbol{E}_2^*). \qquad (28)\end{aligned}$$

The $E$-field energy content of the superposed plane-waves (and, by the same token, that of their $B$-field) is seen to be the sum of the energies of the individual plane-waves. In contrast, the angular momentum within the same cubic volume $(L_x,L_y,L_z) = (\pi/k_x, 1.0, 1.0)$ is given by

$$\begin{aligned}\boldsymbol{\mathcal{J}} &= \int_{-\pi/2k_x}^{\pi/2k_x} x\hat{\boldsymbol{x}}\times\tfrac{1}{2}\varepsilon_0\mathrm{Re}\{[\boldsymbol{E}_1 e^{\mathrm{i}(k_x x + k_z z)} + \boldsymbol{E}_2 e^{\mathrm{i}(-k_x x + k_z z)}]\times[\boldsymbol{B}_1^* e^{-\mathrm{i}(k_x x + k_z z)} + \boldsymbol{B}_2^* e^{-\mathrm{i}(-k_x x + k_z z)}]\}\mathrm{d}x\\ &= \tfrac{1}{2}\varepsilon_0\hat{\boldsymbol{x}}\times\mathrm{Re}\left\{(\boldsymbol{E}_1\times\boldsymbol{B}_2^*)\int_{-\pi/2k_x}^{\pi/2k_x} x e^{2\mathrm{i}k_x x}\mathrm{d}x + (\boldsymbol{E}_2\times\boldsymbol{B}_1^*)\int_{-\pi/2k_x}^{\pi/2k_x} x e^{-2\mathrm{i}k_x x}\mathrm{d}x\right\} \leftarrow \text{integration by parts}\end{aligned}$$

$$= (\pi\varepsilon_0/4k_x^2\omega)\hat{\boldsymbol{x}} \times \text{Im}\{-\boldsymbol{E}_1 \times (\boldsymbol{k}_2 \times \boldsymbol{E}_2^*) + \boldsymbol{E}_2 \times (\boldsymbol{k}_1 \times \boldsymbol{E}_1^*)\}$$

$\boldsymbol{E}_1 \cdot \boldsymbol{k}_1 = \boldsymbol{E}_2 \cdot \boldsymbol{k}_2 = 0$

$$= (\pi\varepsilon_0/4k_x^2\omega)\hat{\boldsymbol{x}} \times \text{Im}\{-(\boldsymbol{E}_1 \cdot \boldsymbol{E}_2^*)\boldsymbol{k}_2 + (\boldsymbol{E}_1 \cdot \boldsymbol{k}_2)\boldsymbol{E}_2^* + (\boldsymbol{E}_2 \cdot \boldsymbol{E}_1^*)\boldsymbol{k}_1 - (\boldsymbol{E}_2 \cdot \boldsymbol{k}_1)\boldsymbol{E}_1^*\}$$

$$= (\pi\varepsilon_0/4k_x^2\omega)\text{Im}\{(\boldsymbol{E}_1 \cdot \boldsymbol{E}_2^* - \boldsymbol{E}_2 \cdot \boldsymbol{E}_1^*)k_z\hat{\boldsymbol{y}} + [\boldsymbol{E}_1 \cdot (\boldsymbol{k}_2 - \boldsymbol{k}_1)](\hat{\boldsymbol{x}} \times \boldsymbol{E}_2^*) - [\boldsymbol{E}_2 \cdot (\boldsymbol{k}_1 - \boldsymbol{k}_2)](\hat{\boldsymbol{x}} \times \boldsymbol{E}_1^*)\}$$

$$= (\pi\varepsilon_0/4k_x^2\omega)\text{Im}\{(\boldsymbol{E}_1 \cdot \boldsymbol{E}_2^* - \boldsymbol{E}_2 \cdot \boldsymbol{E}_1^*)k_z\hat{\boldsymbol{y}} - 2k_x(\boldsymbol{E}_1 \cdot \hat{\boldsymbol{x}})(\hat{\boldsymbol{x}} \times \boldsymbol{E}_2^*) - 2k_x(\boldsymbol{E}_2 \cdot \hat{\boldsymbol{x}})(\hat{\boldsymbol{x}} \times \boldsymbol{E}_1^*)\}$$

adding the real-valued $\boldsymbol{E}_1 \cdot \boldsymbol{E}_1^*$ and $\boldsymbol{E}_2 \cdot \boldsymbol{E}_2^*$ does not alter the imaginary part of the expression

$$= -\left(\frac{\pi\varepsilon_0}{2k_x\omega}\right)\text{Im}\left\{\left(\boldsymbol{E}_1 \cdot \frac{\boldsymbol{E}_1^* - \boldsymbol{E}_2^*}{2k_x} + \boldsymbol{E}_2 \cdot \frac{\boldsymbol{E}_1^* - \boldsymbol{E}_2^*}{2k_x}\right)k_z\hat{\boldsymbol{y}} + (\boldsymbol{E}_1 \cdot \hat{\boldsymbol{x}})(\hat{\boldsymbol{x}} \times \boldsymbol{E}_2^*) + (\boldsymbol{E}_2 \cdot \hat{\boldsymbol{x}})(\hat{\boldsymbol{x}} \times \boldsymbol{E}_1^*)\right\}. \quad (29)$$

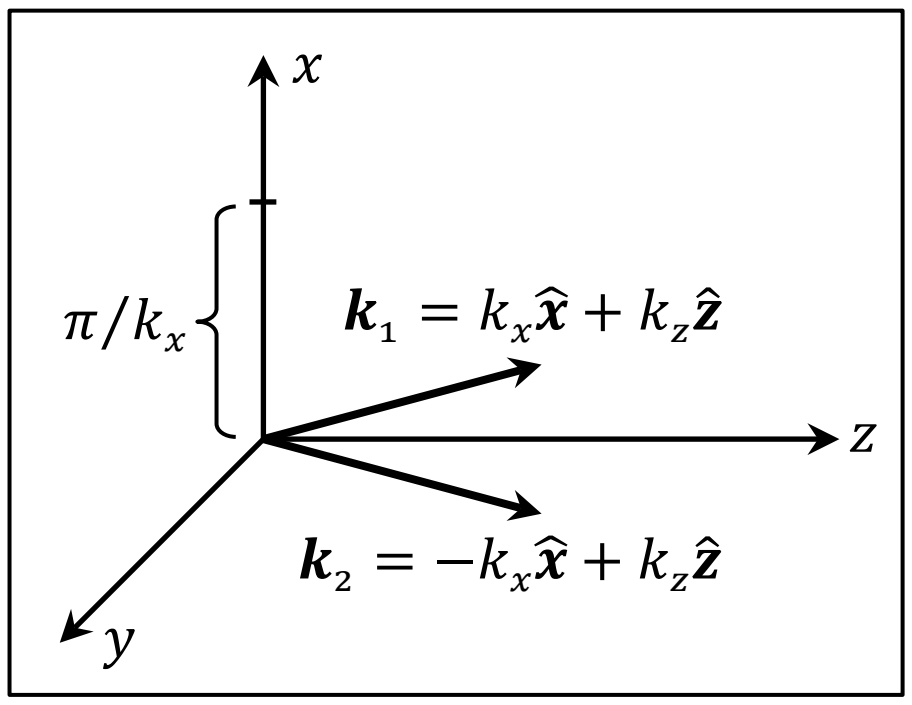

**Fig.3**. Upon interfering in free space, a pair of plane-waves with the same frequency $\omega$ and wave-vectors $\boldsymbol{k}_{1,2} = \pm k_x\hat{\boldsymbol{x}} + k_z\hat{\boldsymbol{z}}$ produce intensity fringes parallel to the $yz$-plane with period $\pi/k_x$ along the $x$-axis.

The first two terms on the right-hand side of Eq.(29) represent the orbital, while the last two terms correspond to the spin angular momentum of the plane-wave pair. Invoking the same line of reasoning as used in the preceding section, we find that each fringe contains exactly one-half of the angular momentum associated with individual EM plane-waves in the continuum limit.[1,3] The trouble with this method of calculation is that it allows each plane-wave to interfere with all the other plane-waves that have the same frequency $\omega$ and, therefore, to contribute to the overall angular momentum in accordance with Eq.(29). At the very minimum, since each plane-wave can have four nearest neighbors of the same frequency in the $k$-space (i.e., neighbors located on the spherical surface of radius $|\boldsymbol{k}| = \omega/c$), its contribution to the overall angular momentum becomes twice as large as the expected contribution by individual plane-waves in the continuum limit.

**5. Resolution of the conundrum**. Equations (15) and (16) show that $\iiint e^{\mathrm{i}(\boldsymbol{k}_1-\boldsymbol{k}_2)\cdot\boldsymbol{r}}\mathrm{d}\boldsymbol{r}$ is an oscillating function of $q = |\boldsymbol{k}_1 - \boldsymbol{k}_2|$, with oscillations that grow ever faster as $R \to \infty$. At the origin of the $k$-space, where $q = 0$, the integral equals $V = 4\pi R^3/3$, which goes to infinity as $R \to \infty$. However, the integral over the entire $k$-space of the function $\iiint e^{\mathrm{i}(\boldsymbol{k}_1-\boldsymbol{k}_2)\cdot\boldsymbol{r}}\mathrm{d}\boldsymbol{r}$ (with $q$ extending from zero to infinity) equals the constant $(2\pi)^3$ (i.e., irrespective of $R$), as seen below.

$\int_0^\infty \cos\zeta\,\mathrm{d}\zeta = 0$

$$\int_{q=0}^{\infty}(4\pi/q^3)[\sin(qR) - qR\cos(qR)](4\pi q^2)\mathrm{d}q = 16\pi^2\int_{\zeta=0}^{\infty}[(\sin\zeta/\zeta) - \cos\zeta]\mathrm{d}\zeta = (2\pi)^3. \quad (30)$$

Thus, in the limit when $R \to \infty$, the function $\iiint e^{\mathrm{i}(\boldsymbol{k}_1-\boldsymbol{k}_2)\cdot\boldsymbol{r}}\mathrm{d}\boldsymbol{r}$ approaches $(2\pi)^3\delta(|\boldsymbol{k}_1 - \boldsymbol{k}_2|)$, which is a three-dimensional delta-function centered at the origin of the $k$-space. Consequently, the cross-terms involving $\boldsymbol{E}_1 \cdot \boldsymbol{E}_2^*$ and $\boldsymbol{E}_2 \cdot \boldsymbol{E}_1^*$ in Eq.(21) should be removed from the expression of $\mathcal{E}_{\text{electric}}$ by dint of the sifting property of the $\delta$-function. In other words, there is no interference between a pair of plane-waves as far as the EM energy of the pair is concerned.

As for the vector function $\iiint \boldsymbol{r}e^{\mathrm{i}(\boldsymbol{k}_1-\boldsymbol{k}_2)\cdot\boldsymbol{r}}\mathrm{d}\boldsymbol{r}$ appearing in the expression of the total angular momentum $\boldsymbol{\mathcal{J}}$ given by Eq.(23), it is related to the $k$-space gradient of the scalar function $\iiint e^{\mathrm{i}(\boldsymbol{k}_1-\boldsymbol{k}_2)\cdot\boldsymbol{r}}\mathrm{d}\boldsymbol{r}$, as follows:

$\hat{\boldsymbol{q}} = (\boldsymbol{k}_1 - \boldsymbol{k}_2)/|\boldsymbol{k}_1 - \boldsymbol{k}_2|$

$$\iiint \boldsymbol{r}e^{\mathrm{i}(\boldsymbol{k}_1-\boldsymbol{k}_2)\cdot\boldsymbol{r}}\mathrm{d}\boldsymbol{r} = -\mathrm{i}\boldsymbol{\nabla}_q \iiint e^{\mathrm{i}(\boldsymbol{k}_1-\boldsymbol{k}_2)\cdot\boldsymbol{r}}\mathrm{d}\boldsymbol{r} = -\mathrm{i}\partial_q\{(4\pi/q^3)[\sin(qR) - qR\cos(qR)]\}\hat{\boldsymbol{q}}$$

$$= -\mathrm{i}4\pi\{-(3/q^4)[\sin(qR) - qR\cos(qR)] + (1/q^3)[\cancel{R\cos(qR)} - \cancel{R\cos(qR)} + qR^2\sin(qR)]\}\hat{\boldsymbol{q}}$$

$$= (\mathrm{i}4\pi/q^4)[3\sin(qR) - 3qR\cos(qR) - (qR)^2\sin(qR)]\hat{\boldsymbol{q}}$$

$$= \frac{\mathrm{i}4\pi[3\sin(|\boldsymbol{k}_1-\boldsymbol{k}_2|R) - 3|\boldsymbol{k}_1-\boldsymbol{k}_2|R\cos(|\boldsymbol{k}_1-\boldsymbol{k}_2|R) - (|\boldsymbol{k}_1-\boldsymbol{k}_2|R)^2\sin(|\boldsymbol{k}_1-\boldsymbol{k}_2|R)](\boldsymbol{k}_1-\boldsymbol{k}_2)}{|\boldsymbol{k}_1-\boldsymbol{k}_2|^5}. \tag{31}$$

As expected, the above result is in complete accord with Eqs.(17) and (19). Thus, in the limit when $R \to \infty$, the vector function $\iiint \boldsymbol{r}e^{\mathrm{i}(\boldsymbol{k}_1-\boldsymbol{k}_2)\cdot\boldsymbol{r}}\mathrm{d}\boldsymbol{r}$ coincides with $-\mathrm{i}(2\pi)^3\boldsymbol{\nabla}_q\delta(|\boldsymbol{k}_1 - \boldsymbol{k}_2|)$. The sifting property of this gradient of a $\delta$-function ensures that only plane-wave pairs that are very close to each other (i.e., immediate neighbors) in the discretized $k$-space contribute to the total angular momentum $\boldsymbol{\mathcal{J}}$ of Eq.(23).

All in all, the unexpected factor of 2 that arose in our analysis of angular momentum in a discrete Cartesian coordinate system, and also the difficulties encountered when the energy and angular momentum were computed within a large spherical volume, can be avoided if transitions to the limit (i.e., $L \to \infty$ or $R \to \infty$) are carried out at the outset. Thus, in Cartesian coordinates, the fundamental properties of the gradient of a $\delta$-function ensure that each plane-wave interferes with only three of its nearest neighbors in the $k$-space — one along each of the $k_x, k_y, k_z$ axes — to produce its contribution to total $\boldsymbol{\mathcal{J}}$. Similarly, the energy and the angular momentum integrated over a large spherical volume simplify substantially as a result of the emergent $\delta$-functions sifting out individual plane-waves (in the case of EM energy) or very close and adjacent pairs of plane-waves (in the case of angular momentum). In light of Ref.[3], one may also conclude that the contributions to the overall angular momentum $\boldsymbol{\mathcal{J}}$ arise from interference between each plane-wave, say, $(\boldsymbol{E}_1, \boldsymbol{k}_1, \omega)$, and only two adjacent plane-waves, $(\boldsymbol{E}_2, \boldsymbol{k}_2, \omega)$ and $(\boldsymbol{E}_3, \boldsymbol{k}_3, \omega)$, when these $k$-vectors are very close to each other and satisfy the relations $|\boldsymbol{k}_1| = |\boldsymbol{k}_2| = |\boldsymbol{k}_3| = \omega/c$ and $(\boldsymbol{k}_1 - \boldsymbol{k}_2) \perp (\boldsymbol{k}_1 - \boldsymbol{k}_3)$.

**6. Concluding remarks**. The energy content of an EM wavepacket in free space can be expressed as a sum (or integral) over the energies of its individual plane-wave constituents. The energy of each plane-wave is equally split between its $\boldsymbol{E}$ and $\boldsymbol{B}$ fields, with the total energy-density (i.e., energy per unit volume) being equal to $\tfrac{1}{2}\varepsilon_0\boldsymbol{E}_0 \cdot \boldsymbol{E}_0^*$, where $\boldsymbol{E}_0$ is the complex-amplitude of the plane-wave's electric field. In quantum optics, the $E$-field amplitude of a plane-wave containing a single photon of frequency $\omega$ within a large spatial volume $V$ is taken to be $\sqrt{\hbar\omega/(2\varepsilon_0 V)}\,\hat{\boldsymbol{e}}$, where $\hbar$ is Planck's reduced constant and the (generally complex-valued) unit-vector $\hat{\boldsymbol{e}}$ is the polarization of the EM mode in which the photon resides.[**] Thus, the energy of the photon — as if it were spread across the entire volume $V$ —comes out as $\hbar\omega$. In general, each mode contains a superposition of number-states $|n\rangle$ and, roughly speaking, the action of the annihilation operator $\hat{a}$ on the number-state $|n\rangle$, namely, $\hat{a}|n\rangle = \sqrt{n}|n-1\rangle$, brings the corresponding $E$-field amplitude to $\sqrt{n\hbar\omega/(2\varepsilon_0 V)}\,\hat{\boldsymbol{e}}$.

[**] In quantum electrodynamics, $\hat{\boldsymbol{E}}(\boldsymbol{r},t) = \mathrm{i}\sqrt{\hbar\omega/(2\varepsilon_0 V)}\,\{\hat{\boldsymbol{e}}\exp[\mathrm{i}(\boldsymbol{k}\cdot\boldsymbol{r} - \omega t)]\,\hat{a} - \hat{\boldsymbol{e}}^*\exp[-\mathrm{i}(\boldsymbol{k}\cdot\boldsymbol{r} - \omega t)]\,\hat{a}^\dagger\}$ is the electric field operator in free space, acting on a propagating plane-wave mode $(\omega, \boldsymbol{k}, \hat{\boldsymbol{e}})$ of the EM field. Each plane-wave is assumed to occupy a large volume $V$ in our three-dimensional space. To relate the above $E$-field operator to the $E$-field of the classical EM plane-wave defined in Eq.(1), one must equate $|\boldsymbol{E}_0|$ with $\sqrt{2\hbar\omega/(\varepsilon_0 V)}$. Note that the $E$-field operator is written here in the Heisenberg picture, where, in contrast to the Schrödinger picture, the operators are functions of time and the states are time-independent.[1,2]

Propagating plane-waves in free space are found to be natural and convenient modes of the classical EM field for transition to quantum optics. In addition to possessing an EM energy of $\hbar\omega$ per photon, each propagating plane-wave mode carries a linear momentum of $(\hbar\omega/c)\hat{\boldsymbol{\kappa}}$ per photon, where $\hat{\boldsymbol{\kappa}}$ is the unit-vector along the direction of $\boldsymbol{k}$.[1-4] A spin angular momentum of $\pm\hbar\hat{\boldsymbol{\kappa}}$ can be similarly associated with each photon that occupies the plane-wave mode $(\omega, \boldsymbol{k}, \hat{\boldsymbol{e}})$ in free space, provided that the mode's polarization unit-vector $\hat{\boldsymbol{e}}$ is taken to be either $\boldsymbol{e}' + \mathrm{i}\boldsymbol{e}''$ or $\boldsymbol{e}' - \mathrm{i}\boldsymbol{e}''$, where $\boldsymbol{e}' \cdot \boldsymbol{e}' = \boldsymbol{e}'' \cdot \boldsymbol{e}'' = ½$ and $\boldsymbol{e}' \cdot \boldsymbol{e}'' = 0$. (The plus and minus signs of $\hat{\boldsymbol{e}} = \boldsymbol{e}' \pm \mathrm{i}\boldsymbol{e}''$ represent, respectively, the right- and left-circular-polarization states.)

In contrast to spin, the orbital angular momentum is a collective property of an entire wavepacket which cannot be apportioned among the individual plane-wave modes that constitute the packet.[1,2] This paper has argued that, in a finely discretized $k$-space, a wavepacket's orbital angular momentum is the sum total of contributions from plane-wave pairs possessing identical frequencies $\omega$ but wave-vectors $\boldsymbol{k}$ that are adjacent neighbors in the $k$-space. A possible method of computing the orbital angular momentum of an EM wavepacket in free space involves discretizing its plane-wave spectrum in the $k$-space over concentric spherical shells (each shell corresponding to a constant frequency $\omega = c|\boldsymbol{k}|$). Discrete $k$-vectors located on the surface of any given spherical shell should then be paired, once with their immediate neighbor in the direction of increasing polar angle $\theta$, and a second time with their immediate neighbor in the direction of increasing azimuthal angle $\varphi$. Adding up the orbital angular momenta of all the plane-wave pairs thus selected will then yield the total orbital angular momentum of the discretized wavepacket.

**Acknowledgement**. The author is grateful to Ewan M. Wright and Miroslav Kolesik for helpful discussions.